\documentclass[article]{aa}
\usepackage{psfig}
\def\teff{\ifmmode T_{\rm eff} \else $T_{\mathrm{eff}}$\fi}

\def\ltsima{$\buildrel<\over\sim$}
\def\lsim{\lower.5ex\hbox{\ltsima}}

\newcommand{\hii}{H~{\sc ii}}
\newcommand{\ha}{\ifmmode {\rm H}\alpha \else H$\alpha$\fi}
\newcommand{\hb}{\ifmmode {\rm H}\beta \else H$\beta$\fi}
\newcommand{\lya}{\ifmmode {\rm Ly}\alpha \else Ly$\alpha$\fi}

\newcommand{\gtapprox}{\raisebox{-0.5ex}{$\,\stackrel{>}{\scriptstyle\sim}\,$}}

\def\msun{\ifmmode M_{\odot} \else M$_{\odot}$\fi}
\def\zsun{\ifmmode Z_{\odot} \else Z$_{\odot}$\fi}
\def\lsun{\ifmmode L_{\odot} \else L$_{\odot}$\fi}

\def\mup{\ifmmode M_{\rm up} \else M$_{\rm up}$\fi}
\def\mlow{\ifmmode M_{\rm low} \else M$_{\rm low}$\fi}

\def\aap{A\&A}
\def\aaps{A\&AS}

\def\apj{ApJ}
\def\apjl{ApJ}
\def\apjs{ApJS}
\def\mnras{MNRAS}

\newcommand{\oh}{\ifmmode 12 + \log({\rm O/H}) \else$12 + \log({\rm
O/H})$\fi}

\newcommand{\oiii}{[O~{\sc iii}]}
\def\Nii{[N\small II]\normalsize $\lambda\lambda$6548,6584}
\def\Sii{[S~{\sc ii}] $\lambda\lambda$6717,6731}

\def\Oii{[O~{\sc ii}] $\lambda$3727}
\def\Oiii{[O~{\sc iii}] $\lambda\lambda$4959,5007}

\begin{document}
\title {Discovery of a faint $R$-band drop-out: a
strongly reddened lensed star forming galaxy at $z=1.68$
\thanks{Based on observations collected with the
ESO VLT-UT1 Antu Telescope (70.A-0355, 271.A-5013),
the Hubble Space Telescope ({\it HST})
and the Canada-France-Hawaii telescope.}
}
\author{ J.~Richard\inst{1},
D.~Schaerer\inst{2,1},
R.~Pell\'o\inst{1},
J.-F.~Le~Borgne\inst{1},
J.-P.~Kneib\inst{3,1}
}
\offprints{J. Richard, jrichard@ast.obs-mip.fr }
\institute{
Laboratoire d'Astrophysique (UMR 5572),
Observatoire Midi-Pyr\'en\'ees,
14 Avenue E. Belin, F-31400 Toulouse, France
\and
Observatoire de Gen\`eve,
51, Ch. des Maillettes, CH-1290 Sauverny, Switzerland
\and
Caltech Astronomy, MC105-24, Pasadena, CA 91125, USA
}
\date{Received 01 oct 2003; accepted 06 nov 2003}
\authorrunning{J. Richard et al.}{ }
\titlerunning{Discovery of a faint $z=1.68$ galaxy}{ }

\abstract{ We report the discovery of an unusual emission line galaxy
at redshift $z=1.68$ found from near-IR imaging and spectroscopic
observations with ISAAC/VLT of a faint gravitationally lensed
$R$-band drop-out behind the lensing cluster Abell 1835.
From the emission lines of \Oiii, and \hb\ this galaxy shows a
moderate to high excitation and line equivalent widths typical
of relatively metal-poor \hii\ galaxies.
Its apparent $J$ magnitude translates to an absolute $B$-band
magnitude $M_B \sim$ --16.4
once corrected for a gravitational magnification of 1.8 magnitudes.
This makes it one of the faintest intermediate redshift galaxies
known so far. \\
From the presence of emission lines and the available $VRIJHK$
photometry we determine constraints on its dominant stellar
population. The only viable fit found is for a
fairly young ($\sim$ 6--9 Myr) burst suffering from a considerable
extinction ($A_V \sim$ 1.2--1.8 mag). We find that this object resembles
strongly \hii\ galaxies and intermediate redshift compact emission line
galaxies, albeit suffering from a larger extinction
than usual. We also discuss the possible contamination introduced by
such $R$-band drop-out galaxies in searches for $z \ga 5$ galaxies.
}

\maketitle


\section{Introduction}

The strong lensing effect due to clusters of galaxies
-- typically yielding a magnification by 1--3 magnitudes --
has extensively been used in the last decade to identify and to probe the
population of galaxies at $z \gtapprox 1$ towards the faint
end of the luminosity function and beyond the limits of
conventional spectroscopic samples.
(see e.g.\ Mehlert et al.\ 2001, Ellis et al. 2001, Pell\'o et al. 2003 and
references therein).

The development of near-IR spectrographs on 10m class
telescopes has allowed the study of the rest-frame optical properties
of galaxies using the same emission lines
all the way from the local universe to $z \sim 4$
(Pettini et al.\ 2001, Erb et al.\ 2003).
The contribution of lensing clusters to these detailed studies is
already significant. In particular, the metallicity--luminosity and the
mass--metallicity relations for intrinsically faint lensed
galaxies, as compared to reference samples at different redshifts,
have recently been studied by Lemoine-Busserolle et al. (2003).
Despite these efforts, the sample of $z \sim$ 1--3 galaxies
observed spectroscopically in the near-IR is still dramatically small.
Furthermore all of these objects were selected from optical
imaging and spectroscopy.

Here we report the discovery of a faint emission line galaxy at
$z\sim1.7$ (named \#2582 hereafter) discovered recently during
near-IR spectroscopic observations targetting $z \sim$ 8 to 10
candidates selected from deep JHK ISAAC imaging in a $2'
\times 2'$ field centered on the lensing cluster Abell 1835
that reached depths of 25.1 in J, 24.3 in H and 24.3 in K (3$\sigma$
on 4 pixels, Vega system). This survey area corresponds to about 2.9
arcmin$^{2}$ at $z=1.7$, after correction of lensing magnification.
Object \#2582 fulfills the following photometric selection criteria: it is an
$R$-band drop-out on HST images, with blue $H-K \le 0.5$ and red $J-H
\ge 0.8$. Because this object was marginally detected 
in $I$, it was considered as an interesting secondary target,
potentially a $z \ga 5$ galaxy or possibly a source of ``contamination''
in searches of high $z$ galaxies. Results on this specific project will be
reported elsewhere (Richard et al., in preparation).
Furthermore, with $R-K \ga 4$ this object is quite red, although
not exactly comparable to extremely red objects (ERO).
Our observations thus reveal that we are dealing with a previously
unknown type of galaxy at intermediate redshift:
a strongly reddened low-luminosity star-forming galaxy.

In Section~\ref{observations} we summarize the observations
used in the present study.
The spectroscopic properties of \#2582 and its nature are
addressed in Sects.\ \ref{results} and \ref{2582} respectively.
Implications from our finding are discussed in Sect.\ \ref{discussion}.

\section{Photometric and Spectroscopic Data}\label{observations}

\begin{figure}[h]
\centerline{
 \begin{tabular}{cc}
 \multicolumn{2}{c}{
 \psfig{figure=Fj011_f1.eps,width=8cm,angle=0}
 }
 \\
 \psfig{figure=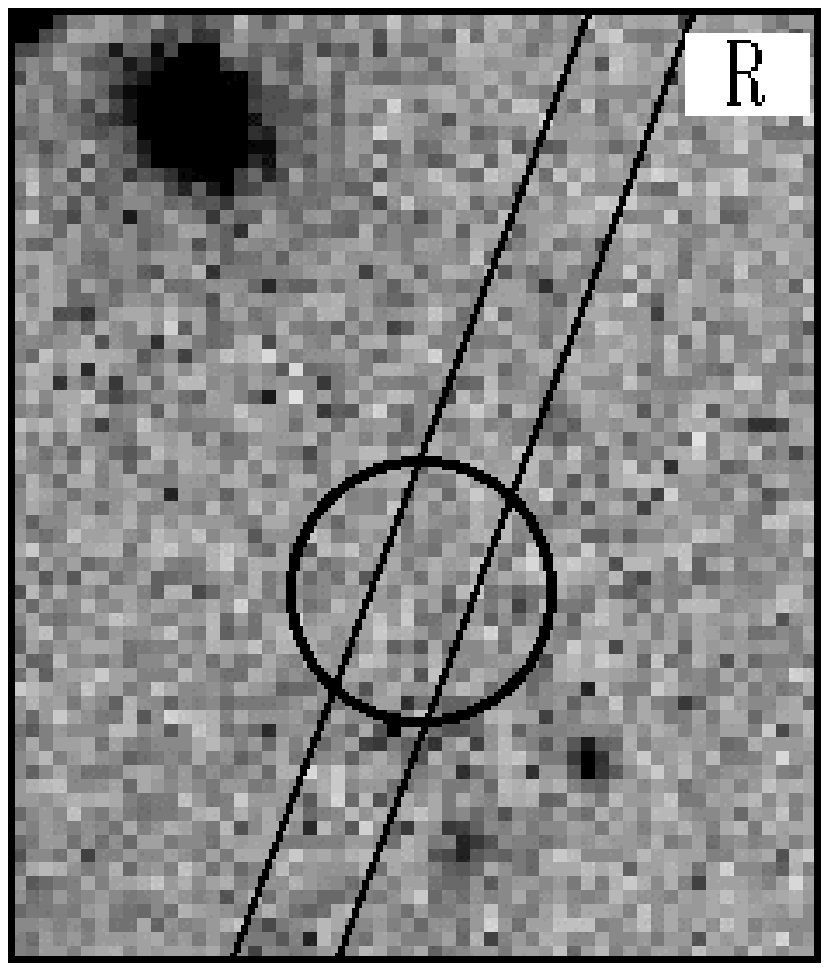,width=3cm,angle=0}
 &
 \psfig{figure=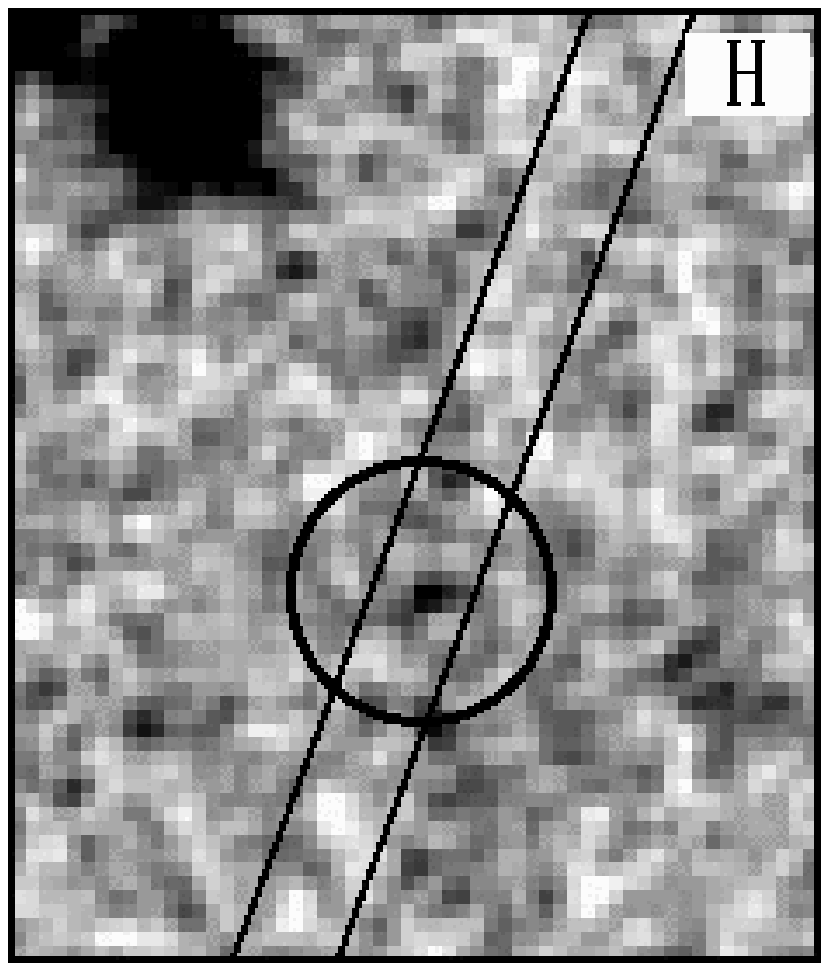,width=3cm,angle=0}
 \\
 \end{tabular}
 }
\caption{Composite $JHKs$ ISAAC image of the cluster Abell 1835 (top),
showing the location of object \#2582 (circle) and the slit
configuration used for spectroscopy with ISAAC.
Superposed are the theoretical critical lines at z=1.6755.
On the bottom, close-ups images in the $R$-band (F702W) {\it HST}/{\it
WFPC2} (left) and the $H$-band ISAAC, smoothed with a sigma $=$ 1
pixel gaussian (right).}
\label{fig_image}
\end{figure}

Ultradeep $JHKs$ images were obtained with ISAAC/VLT
on the central $2 \times 2$ arcmin of Abell 1835 ($z=0.253$),
with ISAAC/VLT in service mode during February 2003.
Photometric data were complemented by deep $VRI$ observations taken
at CFHT, and WFPC2/HST images in the $R$ band
(F702W, Smith et al. 2003).
Photometry was performed using the
{\it SExtractor} package (Bertin \& Arnouts 1996).
A full report of these observations, including particular data
reduction procedures, will be published elsewhere.

During a run between 29 June and 3 July 2003
we have carried out spectroscopic observations in the $J$ band
with ISAAC (SW mode) which included \#2582 as a secondary target aligned
``coincidentally'' in the slit position shown in Fig.\ \ref{fig_image}.
The slit width was $1''$, and the seeing varied between $0.4''$ and $0.5''$.
The coordinates of this object are $\alpha_{2000}$=14:01:00.700,
$\delta_{2000}$=+02:52:09.55.

\begin{table}
\caption{Main properties of images (central wavelength, exposure time, seeing 
and pixel size) and photometry of target \#2582 in the Vega and AB systems, obtained
within $1.5''$ aperture on seeing matched images. Error bars are from
SExtractor. Limiting magnitudes correspond to a detection limit of
3$\sigma$ on 4 pixels.
}
\begin{center}
\begin{tabular}{llccccc}
\hline

Filter & $\lambda_{\rm eff}$ & t$_{\rm exp}$ & $\sigma$ & pix & mag & mag \\
       &   [$\mu$\rm m]             & [ksec] & [$''$] & [$''$] &     [Vega]     &    [AB]     \\
\hline \hline
$V$ (1)    & 0.54  & 3.75 & 0.76 & 0.206& $>$ 28.3     &   $>$ 28.3 \\
$R$ (1)    & 0.66  & 5.4  & 0.69 & 0.206 & $>$ 28.3     &   $>$ 28.5 \\
$I$ (1)    & 0.81  & 4.5  & 0.78 & 0.206 & 26.8 $\pm$ 0.47 &  27.3 \\
$J$ (2)    & 1.26 & 6.48 & 0.65 & 0.148 & 24.7 $\pm$ 0.45 &   25.6 \\
$H$ (2)    & 1.65 & 13.86 & 0.50 & 0.148 & 23.7 $\pm$ 0.16 &   25.1 \\
$Ks$ (2)    & 2.16 & 18.99 & 0.38 & 0.148 & 24.3 $\pm$ 0.30 &   26.2 \\
\hline
\end{tabular}
\end{center}
\begin{center}
\noindent (1)Czoske et al. 2002;
(2) Richard et al.\ in preparation.
\label{tab_phot}
\end{center}
\end{table}

Spectra were obtained in beam-switching 
mode between two positions A and B, following a sequence ABBA 
(see e.g. Cuby et al. 2003\footnote{ISAAC User Manual:
{\tt http://www.eso.org/instruments/isaac/\\
userman/umhtml1121/index.html}},
Lemoine-Busserolle et al. 2003)
Spectroscopic data
where reduced using IRAF procedures and conforming to the ISAAC
Data Reduction Guide 1.5 \footnote{
{\tt http://www.hq.eso.org/instruments/isaac/index.html}}.
The first sky-subtraction was performed by subtracting one frame from
the other in each AB pair. After removing the 50 Hz pickup that
occured during the last night and flat-fielding these frames,
we wavelength-calibrated the
two-dimensional spectra using the atlas of OH lines (Rousselot et al.,
2000).  Finally, we combined each A-B and B-A frames after suitable
shifts and extracted the one-dimensional spectrum.
We used the observed telluric standards to flux-calibrate and correct for
telluric features in the individual spectra, fitting the hot (O and B)
stars with a blackbody curve.

\section{Results}\label{results}

Table~\ref{tab_phot} summarizes the
photometric data obtained on object \#2582,
computed within a $1.5''$ aperture on seeing matched images.
This source is unresolved, and
it is not detected on the WFPC2/HST image 
($\mu_{F702W} \ge$ 23.8 mag/arcsec$^2$ and $R_{F702W} \ge$ 27.3,
2$\sigma$ on 4 HST pixels).
It is undetected in $V$ and $R$ and only marginally
detected in $I$.
This object would {\bf never} have been selected from
our optical data, since it is only marginally detected in the I band.

\begin{figure}
\psfig{file=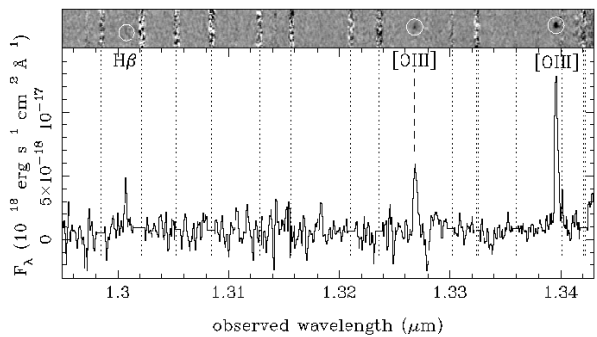,angle=270,width=8.8cm}
\caption{2D and extracted spectra of \#2582 showing the
wavelength interval between [O~{\sc iii}] 5007 and H$\beta$. The
extracted spectrum is flux calibrated and corrected for distorsion.
Dotted vertical lines display the position of the main OH lines.
}
\label{2582_spectrum}
\end{figure}

The spectroscopic observations revealed the presence of 3 emission lines
in 2 overlapping regions of the
$J$ band: 1.285-1.345 $\mu m$ and 1.335-1.395 $\mu m$,
observed with exposure times of 10.8 and 18.9 ksec
respectively. These lines were identified as \Oiii\ and \hb\ at wavelengths
corresponding to an average redshift of $z=1.676$ for this IR-selected
source. The corresponding 2D and extracted spectra are shown in
Fig.~\ref{2582_spectrum}.
As seen on the 2D spectrum, the ``trace'' of the spectrum does not follow a
detector line. Extraction (with iraf task apall) was done by
shifting the fitted ``trace'' of the spectrum of the brighter star used for
slit alignement onto the position of the [OIII]5007\AA\ line of \#2582.
The observed line fluxes are given in Table~\ref{tab_2582}.
Because of the excellent seeing conditions and the slit
width, we can safely consider that the bulk of the flux from this
unresolved source was included in the slit.
The lines are not resolved as compared to the instrumental
profile measured using the OH sky lines. Thus, the line-of-sight velocity
dispersion $\sigma$ should be smaller than 20-30 km/s.

\begin{table}
\caption{Emission lines detected in the
target \#2582, uncorrected for lensing magnification. }
\begin{center}
\begin{tabular}{ccc}
\hline
Line Id. & $z$ & $F_{\lambda}$              \\
         &   & $ 10^{-17}$ erg/s/cm$^2$    \\
\hline \hline
\oiii\ $\lambda$5007 & 1.6755 & 3.93 $\pm$ 0.1 \\
\oiii\ $\lambda$4959 & 1.6757 & 1.39 $\pm$ 0.1 \\
 H$\beta$      & 1.6759 & 0.66 $\pm$ 0.1 \\
\hline
\end{tabular}
\label{tab_2582}
\end{center}
\end{table}

From simple SED fit (cf.\ below) and adopting ``concordance''
cosmological parameters\footnote{
$\Omega_{\rm m}=0.3$, $\Omega_\Lambda = 0.7$, and $H_0 = 70\,{\rm
km\,s^{-1}\,Mpc^{-1}}$},
the apparent $J$ magnitude translates
to an absolute $B$-band magnitude $M_B \sim $ --18.2.
Corrected for the magnification factor of 1.8 mag obtained from
the lensing model by Smith et al.\ (2002), this yields an absolute magnitude
of only $M_B \sim$ --16.4.
Even taking the apparently high extinction into account (cf.\ below)
this still corresponds to $M_B \sim$ --18. to --18.8.
To the best of our knowledge, this makes it the faintest starforming
source at intermediate redshift for which spectroscopic data have been
obtained.
We now discuss the possible nature of this faint galaxy.

\section{The nature of the $z=1.676$ galaxy \#2582}\label{2582}

The observed line ratio \oiii\ $\lambda$5007/\hb\ $\sim$ 5.9
is of moderate to high excitation
typical of relatively metal-poor \hii\ galaxies whose emission
lines are predominantly powered by star formation.
The emission lines are
unresolved, and thus we can exclude a Seyfert 1 galaxy, but in
principle not a Seyfert 2.
However, the \#2582 object is fainter than most Sey 2 (cf.\ Ho et al.\ 1997)
even if corrected for extinction.
With an absolute magnitude of $M_B \sim$ -16.4 (or $\sim$ --18 to
--18.8 after extinction correction) this galaxy
is fainter than Lyman break galaxies at $z \sim 3$ by at least 3 mag,
but similar to the compact narrow emission line galaxies (CNELG)
at $z <$ 1.4 of Guzman et al.\ (1997).
The \hb\ luminosity ($L(\hb) \sim 1.2 \times 10^{41}$ erg s$^{-1}$)
is also comparable to that of CNELG, and to the bright end
of \hii\ galaxies in the local Universe.

The observed emission lines contribute to $\sim$ 56 \% of the observed
$J$-band flux within the slit.
Assuming Case B recombination and zero ($A_V \sim$ 1.8 mag) extinction
a lower limit to the contribution of H$\alpha$ to the H-band is estimated to $\sim$ 17
(30) \%.
We do not expect significant contamination from other emission lines
on the remaining filters.
After correction of the observed broad band flux for the emission lines,
the following
estimate is obtained for the rest-frame \hb\ equivalent width:
$W(\hb)_{\rm rest} \sim$ 139 \AA.
Whereas smaller equivalent widths are typically observed in
large starburst galaxies and in some CNELG, such values are fairly common in low
metallicity \hii\ galaxies (e.g.\ Stasi\'ska \& Izotov 2003).

\begin{figure}
\centerline{\psfig{file=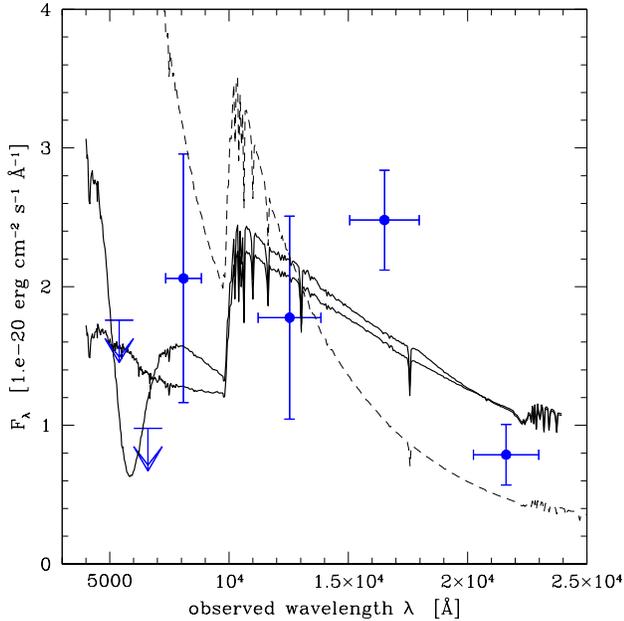,width=8.8cm,angle=0}}
\caption{Observed and modeled broadband SED of \#2582 covering
the $VRIJHK$ bands.
The $J$ and $H$-band fluxes have been corrected for line emission
as indicated in the text. Error bars correspond to 1$\sigma$.
The two best fit spectra using the Seaton (1979) Milky Way extinction
law and the Calzetti et al.\ (2000) attenuation law for starbursts
are shown by solid lines. The corresponding extinction is
$A_V \sim$ 1.2 to 1.8 mag respectively.
The unreddened model SED of a 6 Myr old burst
(taken here for a metallicity of $1/2.5$ solar)
is shown by the dashed line.}
\label{fig_sed}
\end{figure}

The broad-band SED plotted in Fig.\ \ref{fig_sed}
provides further information on the properties of this object.
Basically such a SED can only be reconciled with
a population showing a strong Balmer break and little extinction
or a younger stellar population strongly extinguished in the
rest-frame UV.
This is quantitatively confirmed by SED fits of numerous templates
using the code of Bolzonella et al.\ (2000),
including spectra from the 2001 version of
Bruzual \& Charlot (1993) models, Starburst99
(Leitherer et al.\ 1999), and observed
galaxy templates from Coleman et al.\ (1980) and Kinney et al.\
(1996).
The best fits correspond to burst models with ages $\sim$ 360--510 Myr
and little or no extinction, or bursts of ages $\sim$ 6--9 Myr
with $A_V \sim$ 1.2--1.8 mag depending on the adopted extinction law.
The former explanation is excluded as populations of such age
are not compatible with the presence of emission lines indicative
of young ($\la$ 10 Myr) massive stars.
Furthermore, if present (in quantities sufficient to explain e.g.\
the observed \hb\ flux) the young population will
dominate the rest-frame UV--optical spectrum.
We therefore conclude that the only consistent explanation for
the observed SED of \#2582 is that it is dominated by a
young ($\sim$ 6--9 Myr) population which suffers from a strong
extinction.

This best fit reproduces the observed SED to within $\sim$ 1--2
$\sigma$, as shown in Fig.\ \ref{fig_sed}, using two different
extinction laws: Calzetti et al.\ (2000) for starbursts, with $A_V
\sim$ 1.6--1.8 mag, and the Seaton (1979) Milky Way extinction law
with $A_V \sim$ 1.2--1.4 mag. The later produces a strong ``absorption
bump'' at $\sim$ 5900 \AA.
Thus, we are probably dealing with a
low-metallicity and dusty young starburst.

The best consistency checks of our explanation on the nature of this
source will probably be through \ha\ spectroscopy, in order
to confirm the large extinction and to exclude the Sey 2 possibility.
Deeper optical imaging,
including the $Z$-band, will improve the constraints on the overall SED.
Measurements of other emission lines such as \Oii, \Nii, and \Sii\
will also provide a better understanding of the physical properties and
of the nature of this source.

\section{Discussion and conclusions}\label{discussion}

The $V$ and $R$-band drop-out technique has recently been applied by
various authors in searches of $z \ga 5$ galaxies.
E.g.\ Lehnert \& Bremer (2003) use $RIz$ images taken with the VLT
and apply a $R_{AB}-I_{AB} \ge 1.5$ color criterion to select
galaxies with $z > 4.8$.
Iwata et al.\ (2003) adopt the $V-I_c \ge 2.0$ ``drop-out'' and a
combined $V-I_c$ and $I_c - z^\prime$ criterion to determine
luminosity functions and star formation rates of $z>5$ galaxies
from their SUBARU data.
Would the \#2582 object, a $V$ and $R$-band drop-out, be selected
as a high-$z$ candidate ?

The galaxy \#2582 has $R_{AB}-I_{AB} > 1.2$ and $V-I > 1.5$, close to the
above selection criteria but not formally above the generally adopted
bounds. If deeper optical imaging was available in $V$ or $R$ and $I$,
\#2582 might have been detected in these bands. Furthermore,
the red R-K color makes it a lower priority candidate for
$z>5$ objects selection. However, because of the large
equivalent width of its lines, this class of objects 
is a problem for searches based on narrow-band only
such as Rhoads et al.\ (2003) and for future $z \ge 7$ searches with 
narrow band filters in the J band. 
In order to unambiguously detect $z>5$ sources, a combination of
photometry and spectroscopy, 
similar to the approach of Lehnert \& Bremer (2003) or our own,
is probably to be preferred.

From our observations we conclude that young strongly reddened
starbursts are potential contaminants of
high-z galaxy samples together with stars and low-z elliptical galaxies.
However, from the existing data it is too early to estimate
the fraction of such ``contaminants'' to photometric
samples of $z \ga 5$ candidates.

Finally we may speculate that the classical criteria
used to search for intermediate redshift galaxies probably miss
objects like the one found here quite independently of the
lensing magnification also employed here.
The true number of intermediate $z$ emission line galaxies
observable with near-IR spectrographs is probably larger than
presently thought.

\acknowledgements

We are grateful to T. Contini and M. Lemoine-Busserolle for useful
comments and discussion.
We thank the ESO Director General for a generous allocation of
Director's Discretionary Time for ISAAC spectroscopy (DDT 271.A-5013).
Part of this work was supported by the
French {\it Centre National de la Recherche Scientifique}
and the Swiss National Foundation. JPK acknowledges support from
Caltech and CNRS.

\end{document}